\documentclass[conference]{IEEEtran}
\normalsize
\ifCLASSINFOpdf
\else
\fi
\usepackage{microtype} \DisableLigatures[f]{encoding = *, family = * }
\usepackage{array}
\usepackage{color}
\usepackage{amsfonts}
\usepackage{amssymb}
\usepackage{amsmath}
\usepackage{graphicx}
\usepackage{epstopdf}
\usepackage{cite}
\usepackage{balance}
\usepackage{textcomp}
\usepackage{gensymb}
\usepackage{float}
\usepackage{tabularx}
\usepackage{multirow}
\usepackage{amsthm}
\usepackage{flushend}
\usepackage{subfigure}
\usepackage[T1]{fontenc}
\usepackage{nccmath}
\usepackage{hyperref}
\hypersetup{pdftex,colorlinks=true,allcolors=blue}
\usepackage{hypcap}
\usepackage[margin=0.63in, bottom=0.95in]{geometry}
\usepackage{geometry} 
\newgeometry{left=0.7in,right=0.8in,top=0.7in,bottom=1in}

\newtheorem{lemma}{\bf Lemma}

\makeatletter
  \def\vhrulefill#1{\leavevmode\leaders\hrule\@height#1\hfill \kern\z@}
\makeatother

\begin{document}
\title{\huge The $\alpha$-Lomax Distribution: A Compound  Channel Model}
\author{{Osamah S. Badarneh and Daniel Benevides da Costa}
\thanks{O. S. Badarneh is with the Electrical Engineering Department, School of Electrical Engineering and Information
Technology, German-Jordanian University, Amman 11180, Jordan (e-mail: Osamah.Badarneh@gju.edu.jo).}
\thanks{D. B. da Costa is with the Department of Electrical Engineering, King Fahd University of Petroleum $\&$ Minerals (KFUPM), Dhahran 31261, Saudi Arabia (email: danielbcosta@ieee.org).}
}

\maketitle
\begin{abstract}
In this paper, we propose the $\alpha$-Lomax distribution as a new compound fading channel model. This new distribution generalizes the recently introduced Lomax fading channel model. It is worth noting that the Lomax distribution is a decreasing function, while the $\alpha$-Lomax is a unimodal function, offering greater flexibility in modeling wireless fading channels. In particular, we derive closed-form expressions for the probability density function and cumulative distribution function for the instantaneous signal-to-noise ratio (SNR). Additionally, we provide closed-form expressions for several fundamental  performance metrics, including outage probability, average bit error rate, and channel capacity. Furthermore, we derive closed-form expression for the average block-length error rate in short-packet communications. Moreover, we fit the PDF of the proposed channel model to empirical data obtained from a device-to-device communication system. We also offer simple and accurate approximations for these expressions in the high SNR regime.

 \end{abstract}
\IEEEpeerreviewmaketitle
\section{Introduction}
\IEEEPARstart{T}{he} efficiency and reliability of wireless systems  significantly depend on the characteristics of the wireless channel through which data is transmitted. As such, wireless channel modeling plays a pivotal role in the design, optimization, and performance evaluation of wireless communication systems.

In the literature, many distributions are available that effectively describe the statistics of wireless fading channels. Very recently, the Lomax distribution was proposed as a potential model for characterizing wireless fading channels \cite{10309028,10356745}. To achieve this goal, the authors redefined the Lomax parameters for channel modeling. Subsequently, they provided closed-form expressions for its fundamental statistics, which were then employed to assess the performance of wireless communication systems. In \cite{Fisher}, the Fisher–Snedecor $\cal{F}$ distribution was introduced to model composite fading channels, where the root-mean-square power fluctuation of a Nakagami-$m$ signal is assumed to be influenced by an inverse Nakagami-$m$ random variable (RV). In \cite{Yacoub2002}, the authors introduced the $\alpha$-$\mu$ fading distribution, which generalizes the Nakagami-$m$ fading model by considering the nonlinearity of the propagation medium. In the literature, the Rayleigh fading distribution was used to model the conditions of strong scattering \cite{1622403}. Compound distributions, which involve two distributions, were introduced to offer distributions with more realistic tails. One of these distributions is the $\cal{K}$ distribution \cite{Abraham}, in which the envelope fluctuations of the Rayleigh distribution follow a gamma random variable (RV), characterizing shadowing effects \cite{778479}.

Despite the fact that the Lomax distribution is a decreasing function, it finds utility in various wireless applications \cite{10309028,10356745}, including ultra-reliable and low-latency communications (URLLC), content delivery within device-to-device (D2D) communications, enabling cooperative spectrum sensing in cognitive radio networks, and assessing bit error performance in the presence of interference.

Motivated by the above, in this paper, we propose a generalization of the Lomax distribution, which we call the $\alpha$-Lomax distribution. In this distribution, the reciprocal of the variance in the Rayleigh distribution follows a Gamma RV, and the resulting signal power is obtained as the modulus raised to a certain given power, represented by the parameter $\alpha$. Furthermore, the $\alpha$-Lomax distribution can take the form of either a decreasing function or a unimodal function, depending on its parameters. As a result, it offers greater flexibility for modeling wireless fading channels when compared to its counterpart, the Lomax distribution. The contributions of this paper can be summarized as follows:

\begin{itemize}
  \item Closed-form expressions for the probability density function (PDF) and cumulative distribution function (CDF) of the instantaneous signal-to-noise ratio (SNR) are derived.
  \item Closed-form expressions of some key performance metrics, such as outage probability, average bit error rate (BER), and average channel capacity, are attained. Additionally, a closed-form expression for the average block error rate (BLER) in short-packet communications is derived.
  \item Tight approximations at high SNR regime are derived for the achieved performance metrics.
  \item A physical model for generating samples of the $\alpha$-Lomax distribution is introduced.
\end{itemize}

The rest of this paper is structured as follows: In Section \ref{sec2}, it is provided a physical description of the Lomax fading distribution and then introduce the $\alpha$-Lomax fading model. Additionally, it is derived the PDF and CDF of the instantaneous SNR as well as some statistical metrics, including generalized moment generating function (MGF), and $n$-th moment. Closed-form expressions for the outage probability, average BER, and average channel capacity are derived in Section \ref{sec3}. Furthermore, simple and accurate approximations at high SNR regime are attained. In Section \ref{sec4}, we present results and discussions to validate our analysis. Finally, Section \ref{sec5} concludes this work.

\section{The $\alpha$-Lomax Fading  Distribution}\label{sec2}
The works in \cite{10309028} and \cite{10356745} lack  details regarding the physical generation of the Lomax distribution. Therefore, in this section, we first present a physical model for the aforementioned distribution, and then we define the $\alpha$-Lomax fading distribution. The Lomax fading distribution arises when the reciprocal of the variance in the Rayleigh distribution follows a Gamma random variable (RV). Thus, one can define the signal power, $P$, of the Lomax distribution as follows{\footnote{Unlike the $\alpha$-Lomax, in \cite{Fisher,9096603}, The fluctuations of the envelope of the Nakagami-$m$ distribution, when the variance is fixed, follow an inverse Nakagami RV.}}:
\begin{equation}\label{plomax}
P=R^{2}=  X^{2}+\,Y^{2},
\end{equation}
where $R$ is the signal envelope, $X$ and $Y$ are mutually independent Gaussian processes, with mean $\mathbb{E}[X]=\mathbb{E}[Y]=0$ and variance $\mathbb{E}[X^{2}]=\mathbb{E}[Y^{2}]={1\over2\tau}$, where $\mathbb{E}[\cdot]$ denotes expectation and $\tau$ follows the Gamma distribution whose PDF is given by
\begin{equation}\label{gamdis}
f_{\tau}(\tau) = {\beta^\lambda\over\Gamma(\lambda)}\,\tau^{\lambda-1}\exp{(-\beta\,\tau)}, \quad \tau>0,
\end{equation}
where $\beta>0$ and $\lambda >0$ are, respectively, the scale and shape parameters that determine the severity of the fading channel. $\Gamma(\cdot)$ is the gamma function \cite[Eq. (8.310.1)]{i:ryz}.

\begin{lemma}
 The PDF and CDF of the signal power $Z$ for the $\alpha$-Lomax distribution are given as follows, respectively:
  \begin{equation}\label{envaekn}
f_{Z}(z) =  {\alpha\,\lambda\,\zeta}\,z^{\alpha-1}\left(1+{\zeta\,z^{\alpha}}\right)^{-(\lambda+1)},
\end{equation}
and
 \begin{equation}\label{cdfplomax}
F_{Z}(z) =  1-\left(1+\zeta\,z^{\alpha}\right)^{-\lambda},
\end{equation}
where $\alpha>0$ represents non-linearity of the propagating medium and $\zeta=\left({\Gamma(1+{1\over\alpha})\Gamma(\lambda-{1\over\alpha})\over\Gamma(\lambda)}\right)^{\alpha}$ and $\lambda>{1\over\alpha}$.
\end{lemma}
\begin{IEEEproof}
See the Appendix.
\end{IEEEproof}

\begin{lemma}
  The PDF and CDF of the instantaneous SNR for the $\alpha$-Lomax distribution are given as follows, respectively:
  \begin{equation}\label{snrpdf}
f_{\Gamma}(\gamma) =  {\alpha\lambda\zeta\over\overline{\gamma}^{\alpha}}\,\gamma^{\alpha-1}\left(1+{\zeta\over\overline{\gamma}^{\alpha}}\,\gamma^{\alpha}\right)^{-(\lambda+1)}
\end{equation}
and
\begin{equation}\label{snrcdf}
F_{\Gamma}(\gamma) =  1- \left(1+{\zeta\over\overline{\gamma}^{\alpha}}\,\gamma^{\alpha}\right)^{-\lambda},
\end{equation}
where $\overline{\gamma}=\mathbb{E}[\Gamma]={P_{t}\over N_{0}}$ denotes the average SNR, $P_{t}$ is the transmit power, and $N_{0}$ is the noise power.
\end{lemma}
\begin{IEEEproof}
See the Appendix.
\end{IEEEproof}

To the best of the authors' knowledge, \eqref{snrpdf} and \eqref{snrcdf} represent new findings. It is important to note that the PDF in \eqref{snrpdf} is unimodal when $\alpha>1$ and a decreasing function when  $0<\alpha\leq1$. Additionally, it's worth mentioning that when $\alpha=1$, \eqref{snrpdf} and \eqref{snrcdf} exactly coincide with \cite[Eq. (4)]{10356745} and \cite[Eq. (5)]{10356745}, respectively. Furthermore, since the Lomax distribution, when $\alpha=1$, is a decreasing function, it is expected that the $\alpha$-Lomax distribution can provide a better description of the fading channel.

\begin{lemma}
 The generalized MGF of the instantaneous SNR can be expressed as follows:
 \begin{align}\label{gmgf1}
&\!\!\!\! \mathcal{M}_{\Gamma}^{n}(s) = {\alpha\over s^{n}\Gamma(\lambda)}{\sf H}^{1,2}_{2,1}\!\left[{\zeta\over\overline{\gamma}^{\alpha}\,s^{\alpha}}\left\vert \begin{matrix} (1-\lambda,1),(1-n,\alpha)\\[0.05cm](1,1)\end{matrix}\right.\right],&
\end{align}
where ${\sf H}^{\cdot,\cdot}_{\cdot,\cdot}[\cdot]$ denotes the Fox H-function \cite[Eq. (8.3.1.1)]{a:pru}.
\end{lemma}
\begin{IEEEproof}
See the Appendix.
\end{IEEEproof}

\begin{lemma}
The $n$-th moment of the instantaneous SNR is given by
\begin{equation}\label{meansnr}
\mathbb{E}[\Gamma^{n}] =  \overline{\gamma}^{n}\,{\lambda\over\zeta^{n\over\alpha}}\,{\mathfrak B}\!\left(1+{n\over\alpha},\lambda-{n\over\alpha}\right),
\end{equation}
where ${\mathfrak B}(\cdot,\cdot)$ denotes the Beta function \cite[Eq. (8.384.1)]{i:ryz}.
\end{lemma}
\begin{IEEEproof}
See the Appendix.
\end{IEEEproof}
\section{Performance Analysis}\label{sec3}
\subsection{Outage Probability Analysis}
The outage probability (OP) can be defined as the probability of the data rate of a communication link falls below a specified rate $R_{0}$ [bps/Hz], i.e., $\mathbb{P}_{\text{out}}\triangleq\Pr(\gamma\leq\gamma_{0})=F_{\Upsilon}(\gamma_{0})$, where $\Pr(\cdot)$ denotes probability and  $\gamma_{0}=2^{R_{0}}-1$. Using \eqref{snrcdf}, the OP can be readily obtained as
\begin{equation}\label{outsnr}
\mathbb{P}_{\text{out}} =  1- \left(1+{\zeta\over\overline{\gamma}^{\alpha}}\,\gamma_{0}^{\alpha}\right)^{-\lambda}.
\end{equation}

At high SNR values, i.e., when $\overline{\gamma}\rightarrow\infty$, the OP has the form of $\mathbb{P}_{\text{out}}\approx(G_{c}\overline{\gamma})^{-G_{d}}$, where $G_{c}$ and $G_{d}$ are, respectively, the coding and diversity gains. Thus, using \eqref{outsnr}, a simple approximation for the OP can be obtained as
\begin{align}\label{out}
\mathbb{P}_{\text{out}}^{\infty}\approx\left({1\over\gamma_{0}\,\zeta^{{1\over\alpha}}\,\lambda^{{1\over\alpha}}}\,\overline{\gamma}\right)^{-\alpha}.
\end{align}
Based on \eqref{out}, $G_{c}={1\over\gamma_{0}\,\zeta^{{1\over\alpha}}\,\lambda^{{1\over\alpha}}}$ and $G_{d}=\alpha$. Note that the system's diversity gain depends only on $\alpha$, which means that increasing $\alpha$ will result in a significant change in performance.

\subsection{Bit Error Rate Analysis}
Assuming coherent binary modulations, the average BER $\overline{P_{b}}$, can be obtained using 
\begin{align}\label{ser}
  \overline{P_{b}}={1\over\pi}\int_{0}^{{\pi\over2}}\mathcal{M}_{\Gamma}\!\left({\varphi\over\sin^{2}\theta}\right){\mathrm{d}}\theta,
\end{align}
where $\varphi$ is modulation-dependent parameter. For coherent binary phase-shift keying (BPSK), $\varphi=1$, whereas for coherent binary frequency shift keying (BFSK), $\varphi=0.5$, and for coherent detection of minimum shift keying (MSK), also known as BFSK with minimum correlation, $\varphi=0.715$.  Setting $n=0$ in \eqref{gmgf1} and substituting the result in \eqref{ser}, it yields
\begin{align}\label{ser1}
& \overline{P_{b}}={\alpha\over\pi\Gamma(\lambda)}\int_{0}^{{\pi\over2}}{\sf H}^{1,2}_{2,1}\left[{\zeta\sin^{2\alpha}\theta\over\overline{\gamma}^{\alpha}\,\varphi^{\alpha}}\left\vert \begin{matrix} (1-\lambda,1),(1,\alpha)\\[0.05cm](1,1)\end{matrix}\right.\right]
{\mathrm{d}}\theta.&
\end{align}
Using the change of variable $x=\sin^{2}\theta$ and applying \cite[Eq. (2.25.2.2)]{a:pru}, a closed-form expression for the average BER  can be derived as
\begin{align}\label{ser2}
\overline{P_{b}}={\alpha\over2\pi\Gamma(\lambda)}
{\sf H}^{1,4}_{4,2}\left[{\zeta\over\overline{\gamma}^{\alpha}\,\varphi^{\alpha}}\left\vert \begin{matrix} ({1\over2},\alpha),({1\over2},0),(1-\lambda,1),(1,\alpha)\\[0.05cm](1,1),(0,\alpha)\end{matrix}\right.\right].
\end{align}

At high SNR values, the average BER can be approximated by $\overline{P_{b}}\approx(G_{c}\overline{\gamma})^{-G_{d}}$. Thus, applying \cite[Eq. (1.8.4)]{kilbas} to \eqref{ser2},  the average BER can be simply approximated by
\begin{align}\label{asyser2}
  \overline{P_{b}}\approx
  \left(\left({2\,\varphi^{\alpha}\,\sqrt{\pi}\over\lambda\,\zeta\,\Gamma\left({1\over2}+\alpha\right)}\right)^{{1\over\alpha}}\overline{\gamma}\right)^{-\alpha}.
\end{align}
Hence, $G_{c}=\left({2\,\varphi^{\alpha}\,\sqrt{\pi}\over\lambda\,\zeta\,\Gamma\left({1\over2}+\alpha\right)}\right)^{{1\over\alpha}}$ and confirms that the system's diversity gain is  $G_{d}=\alpha$, as obtained in the outage probability analysis.

\subsection{Average Channel Capacity Analysis}
The average channel capacity, in [bps/Hz], can be found using
 \begin{align}\label{defcap}
 \overline{C}={1\over\ln(2)}\int_{0}^{\infty}{\ln(1+ \gamma)}\,{f}_{\Gamma}(\gamma)\rm{d}\gamma.
 \end{align}
 Substituting \eqref{snrpdf} into \eqref{defcap}, applying respectively \cite[Eq. (8.4.6.5)]{a:pru}, \cite[Eq. (8.4.2.5)]{a:pru}, \cite[Eq. (2.25.1.1)]{a:pru}, and \cite[Eq. (8.3.2.8)]{a:pru}, the following expression is obtained for the average channel capacity
\begin{align}\label{defcap1}
&\overline{C}={\alpha\over\ln(2)\Gamma(\lambda)}
{\sf H}_{3,3}^{3,2}\left[{\zeta\over\overline{\gamma}^{\alpha}}\left\vert \begin{matrix} (1-\lambda,1),(0,\alpha),(1,\alpha)\\[0.05cm](1,1),(0,\alpha),(0,\alpha)\end{matrix}\right.\right].&
\end{align}

When $\overline{\gamma}\rightarrow\infty$, the average channel capacity can be approximated by
\begin{align}\label{asycap1}
	\overline{C}\approx {1\over\ln(2)}\int_{0}^{\infty}{\ln(\gamma)}\,{f}_{\Gamma}(\gamma)\rm{d}\gamma.
\end{align}

Substituting \eqref{snrpdf} into \eqref{asycap1} and using \cite[Eq. (2.6.4.7)]{pruv1}, the approximated average channel capacity can be expressed as
\begin{align}\label{asycap11}
\!\!\!	\overline{C}\approx {1\over\alpha\ln(2)}\bigg[\ln\left( {\overline{\gamma}^{\,\alpha}\over \zeta}\right) -\gamma_{\rm{E}}  - \Psi(\lambda) \bigg],\!
\end{align}
where $\gamma_{\rm{E}}$ \cite[Eq. (8.367.1)]{i:ryz} and $\Psi(\cdot)$ \cite[Eq. (8.360.1)]{i:ryz}  are the Euler's constant and digamma function, respectively.

\section{Short-Packet Communications}
Short-packet, i.e., finite block-length,  can support ultra-reliable communications. For such communications, the BLER can be accurately approximated by
\begin{align}
 {\sf BLER}\simeq \left\lbrace \begin{array}{ll}1,& \gamma\leq \mu\\ \frac{1}{2}- \frac{\delta }{\sqrt{2\pi }}(\gamma-\eta),&\mu < \gamma < \upsilon\\ 0,& \gamma\geq \upsilon \end{array} \right. \,
 \end{align}
where $\mu=\eta-\sqrt{{\pi\over2\delta^2}}$, $\upsilon=\eta+\sqrt{{\pi\over2\delta^2}}$, $\eta=2^{{K\over N}}-1$,  $\delta=\sqrt{{N\over2\pi}}\left(2^{{2K\over N}}-1\right)^{-{1\over2}}$, $N$ and $K$ are respectively represent the block-length and the number of information bits in each finite block-length. Therefore, over fading channels, the average BLER can be expressed as
\begin{align}\label{bler}
  &\overline{{\sf BLER}}\simeq\int_{0}^{\mu}f_{\Gamma}(\gamma)\rm{d}\gamma-{\delta\over\sqrt{2\pi}}\int_{\mu}^{\upsilon}\gamma\,f_{\Gamma}(\gamma)\rm{d}\gamma\cr&+
  \left({1\over2}+{\delta\eta\over\sqrt{2\pi}}\right)\int_{\mu}^{\upsilon}\,f_{\Gamma}(\gamma)\rm{d}\gamma.&
\end{align}
The first integral represents the CDF of $\Gamma$ evaluated at $\mu$. Now, when substituting \eqref{snrpdf} into \eqref{bler}, then an integral of the following form has to be solved. That is,
\begin{align}\label{bler1}
    \mathcal{I} = \int_{\mu}^{\upsilon} \,\gamma^{p+\alpha-1}\left(1+{\zeta\over\overline{\gamma}^{\alpha}}\,\gamma^{\alpha}\right)^{-(\lambda+1)}\rm{d}\gamma,
\end{align}
where $p\in\{0,1\}$. To the best of authors' knowledge, no solution to this integral is available in the literature. To this end, we represent the quantity $\left(1+{\zeta\over\overline{\gamma}^{\alpha}}\,\gamma^{\alpha}\right)^{-(\lambda+1)}$ in terms of Meijer G-function using \cite[Eq. (8.4.2.5)]{a:pru} and then applying the definition of the Meijer G-function \cite[Eq. (8.2.1.1)]{a:pru}. Thus, \eqref{bler1} becomes
\begin{align}\label{bler2}
   & \mathcal{I} = {1\over\Gamma(\lambda+1)}{1\over\mathbb{J}2\pi}\oint_{\mathcal{L}}\left({\zeta\over\overline{\gamma}^{\alpha}}\right)^{s}\Gamma(-s)\Gamma(1+\lambda+s)
   \cr&\times\int_{\mu}^{\upsilon} \,\gamma^{p+\alpha+\alpha s-1} \,\rm{d}\gamma\,\rm{d}s.&
\end{align}
The inner integral w.r.t. $\gamma$ can be solved as
\begin{align}\label{bler3}
   \int_{\mu}^{\upsilon} \,\gamma^{p+\alpha+\alpha s-1} \,{\rm{d}}\gamma={\Gamma\left({p+\alpha\over\alpha}+s\right)(\upsilon^{p+\alpha+\alpha s} - \mu^{p+\alpha+\alpha s})\over\alpha\,\Gamma\left(1+{p+\alpha\over\alpha}+s\right)}.
\end{align}
Substituting \eqref{bler3} into \eqref{bler2}, representing the result in terms of the Meijer G-function \cite[Eq. (8.2.1.1)]{a:pru}, and applying \cite[Eq. (8.4.49.13)]{a:pru}, the final expression for the average BLER can be obtained as
 \begin{align}\label{blerf}
 &\overline{{\sf BLER}}\simeq F_{\Gamma}(\mu)-c_{1}\!\left\{\upsilon^{\alpha+1}\Theta^{(1)}\!\left({\zeta\upsilon^{\alpha}\over\overline{\gamma}^{\alpha}}\right)
\!-\mu^{\alpha+1} \Theta^{(1)}\!\left({\zeta\mu^{\alpha}\over\overline{\gamma}^{\alpha}}\right)\right\}\cr&
-c_{2}\!\left\{\upsilon^{\alpha}\,\Theta^{(0)}\!\left({\zeta\upsilon^{\alpha}\over\overline{\gamma}^{\alpha}}\right)-\mu^{\alpha}\,\Theta^{(0)}\!    \left({\zeta\mu^{\alpha}\over\overline{\gamma}^{\alpha}}\right)\right\},
 \end{align}
 where $c_{1}={\delta\alpha\lambda\zeta\over\sqrt{2\pi}(1+\alpha)\overline{\gamma}^{\alpha}}$, $c_{2}=\left({1\over2}+{\delta\eta\over\sqrt{2\pi}}\right){\lambda\zeta\over \overline{\gamma}^{\alpha}}$, and $\Theta^{(p)}(x)={}_2F_1(1+\lambda;{p+\alpha\over\alpha};{p+2\alpha\over\alpha};-x)$, with ${}_2F_1(\cdot)$ being the hypergeometric function \cite[Eq. (9.14.2)]{i:ryz}.
 At high SNR values, the asymptotic average BLER can be accurately obtained in a simple form as
 \begin{align}\label{blerf}
 &\overline{{\sf BLER}}\approx {\zeta\lambda\mu^{\alpha}\over\overline{\gamma}^{\alpha}}-c_{1}\left\{\upsilon^{\alpha+1}-\mu^{\alpha+1}\right\}
-c_{2}\left\{\upsilon^{\alpha}-\mu^{\alpha}\right\}.
 \end{align}
 Clearly, the system diversity gain depends only on $\alpha$.

\section{Results and Discussions}\label{sec4}
This section is dedicated to validating the mathematical derivations, specifically, the derived analytical expressions for the PDF of the instantaneous SNR, the outage probability, average BER, average channel capacity, and average BELR for short-packet communications. These expressions are evaluated and compared with Monte-Carlo simulations. Most importantly, the discussions in this section offer deeper insights into the obtained results.

\begin{figure}[t!] \centering

         \includegraphics[width=\columnwidth,keepaspectratio]{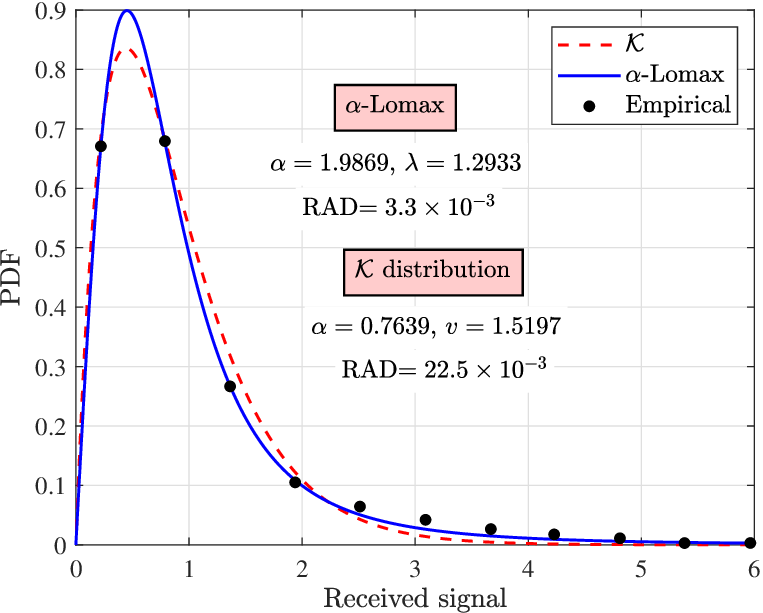}%
       \vspace*{-3mm}
       \caption{Outdoor NLoS empirical and theoretical PDFs fitted to D2D communication systems \cite[Fig. 12]{8270655n}.}\label{pdffit}
\end{figure}

As a practical application, the PDF of the $\alpha$-Lomax is fitted to empirical data obtained from D2D communication systems, as shown in Fig. \ref{pdffit}. Additionally, the PDF of the $\cal{K}$ distribution \cite[Eq. (43)]{Abraham} is also fitted. The $\cal{K}$ distribution is chosen because  it is a compound distribution, similar to the $\alpha$-Lomax. To assess the quality of fit between the theoretical PDFs and the empirical PDF, we calculate the Resistor-Average Distance (RAD) \cite{johnsonsymmetrizing}. The results show that the RAD values for the $\alpha$-Lomax and $\cal{K}$ distributions are respectively  ${\text {RAD}} =  3.3\times10^{-3}$ and ${\text {RAD}} =  22.5\times10^{-3}$, indicating that the $\alpha$-Lomax offers a better fit to the empirical data compared to the $\cal{K}$ distribution.

\begin{figure}[t!] \centering

         \includegraphics[width=\columnwidth,keepaspectratio]{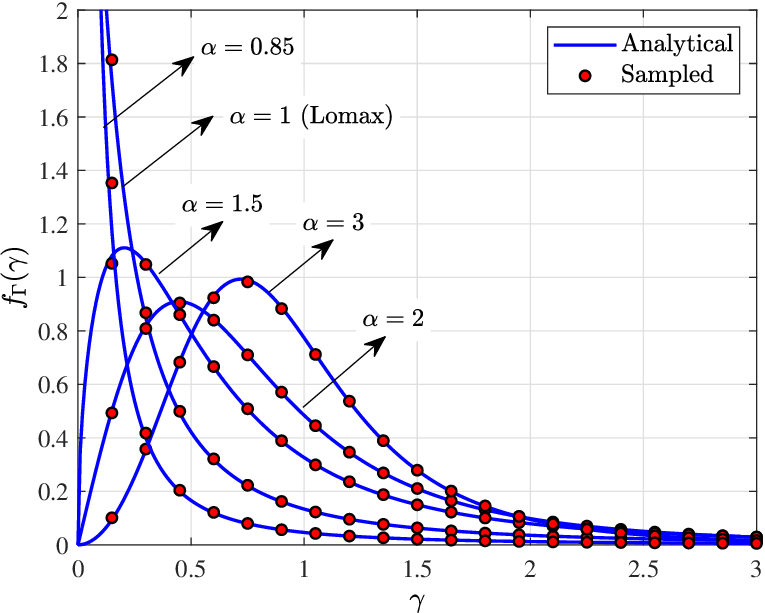}%
        \vspace*{-3mm}
       \caption{$\alpha$-Lomax SNR PDF under varying $\alpha$ with $\lambda=1.25$.}\label{ev1}
\end{figure}

The SNR PDF is plotted in Fig. \ref{ev1} for different values of $\alpha$, by setting $\lambda=1.25$. Clearly, when $\alpha\leq1$, the SNR PDF is a decreasing function, while for $\alpha>1$, the SNR PDF is a unimodal function. Moreover, as $\alpha$ increases, the SNR PDF moves toward the right, indicating better fading conditions. Additionally, the analytical curves of the SNR PDF are in perfect agreement with Monte-Carlo simulation results, validating our analysis and the physical model of the $\alpha$-Lomax relationship.

\begin{figure}[t!] \centering
         \includegraphics[width=\columnwidth,keepaspectratio]{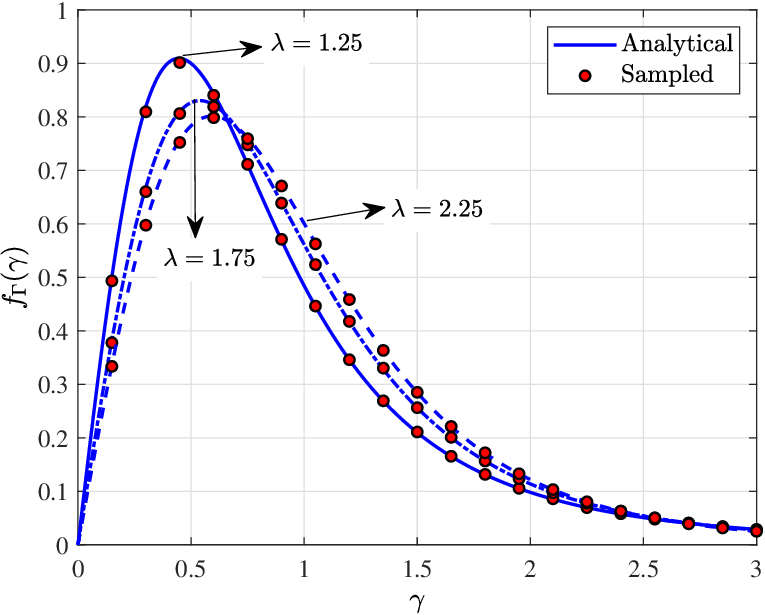}%
       \vspace*{-3mm}
       \caption{$\alpha$-Lomax SNR PDF under varying $\lambda$ with $\alpha=2$.}\label{ev2}
\end{figure}

\begin{figure}[t!] \centering
         \includegraphics[width=\columnwidth,keepaspectratio]{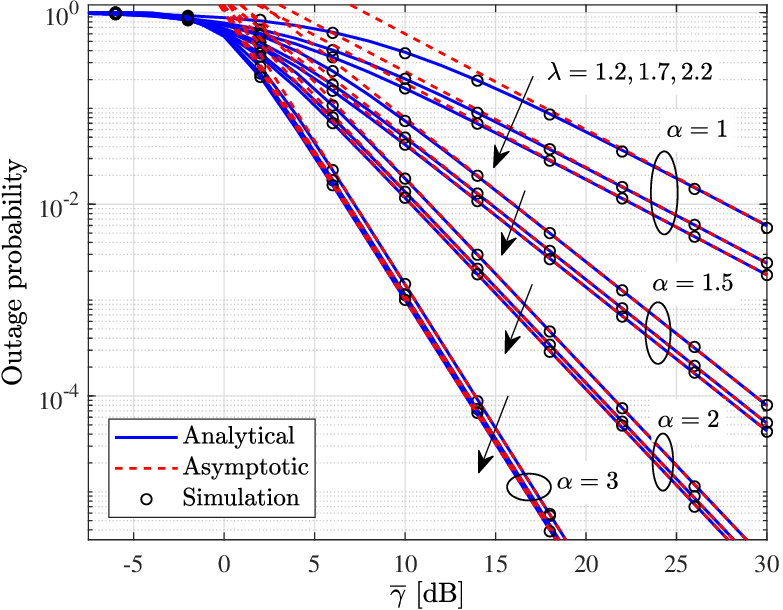}%
       \vspace*{-3mm}
       \caption{Outage performance for different values of $\alpha$ and $\lambda$. $R_{0}=1$ [bps/Hz].}\label{outal}
\end{figure}

The impact of $\lambda$ on the SNR PDF is depicted in Fig. \ref{ev2}. It can be observed that as $\lambda$ increases, the SNR PDF slightly shifts to the right, compared to the effect of $\alpha$, as shown earlier. This observation indicates a slight improvement in channel conditions. Next, we demonstrate that the impact of $\alpha$ on the system performance is more pronounced than that of $\lambda$.
Fig. \ref{outal} illustrates the influence of both parameters, $\alpha$ and $\lambda$, on the outage probability. The results indicate that the outage performance improves as $\alpha$ and(or) $\lambda$ increases. Furthermore, it is noteworthy that increasing $\alpha$ has a more pronounced impact when compared to $\lambda$, whereas the effect of $\lambda$ becomes insignificant at higher values of $\alpha$. Additionally, the asymptotic curves closely match the analytical curves, particularly at high SNR values. Moreover, the slope of the asymptotic curves changes only as $\alpha$ varies, remaining constant with $\lambda$, thus confirming our finding in Eq. \eqref{out}.

\begin{figure}[t!] \centering
         \includegraphics[width=\columnwidth,keepaspectratio]{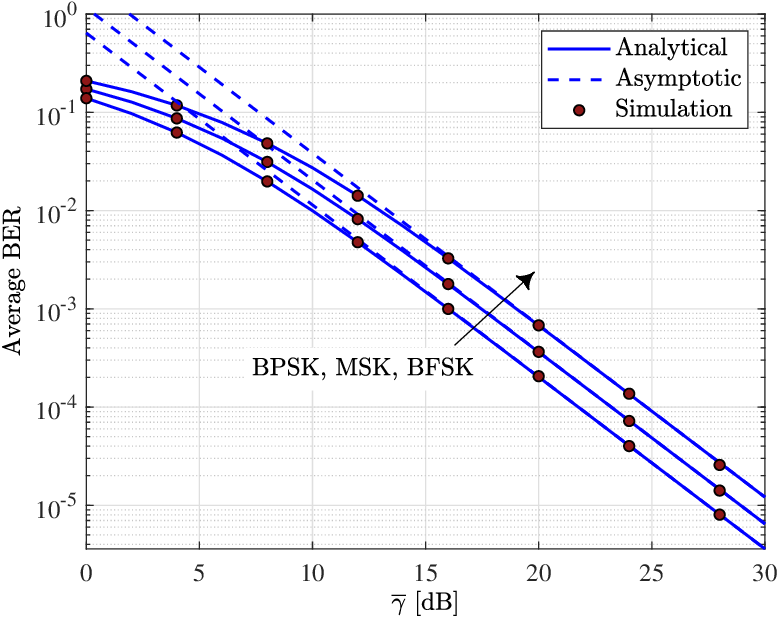}%
       \vspace*{-3mm}
       \caption{Average BER performance for various modulation schemes with $\alpha=1.75$ and $\lambda=1.25$.}\label{berfig}
\end{figure}

 Fig. \ref{berfig} illustrates the average BER across different coherent binary modulation schemes. These findings confirm the accuracy of the derived analytical BER expression and the validity of the asymptotic analysis. Additionally, the results indicate that the diversity gain (i.e., the slope of the asymptotic curves) is independent of the type of modulation.

\begin{figure}[t] \centering
         \includegraphics[width=\columnwidth,keepaspectratio]{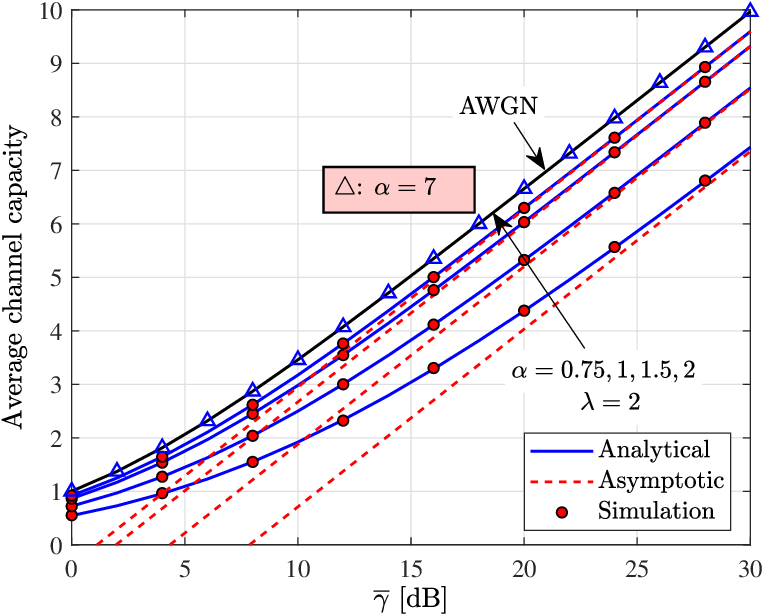}%
       \vspace*{-3mm}
       \caption{Average channel capacity for different values of $\alpha$.}\label{capac1}
\end{figure}

Fig. \ref{capac1} illustrates the variation in the average channel capacity of the $\alpha$-Lomax fading channel across different values of $\alpha$. We also include the capacity of the AWGN channel as a reference point. Notably, as $\alpha$ grows, the average channel capacity experiences enhancement, eventually converging to that of the AWGN channel, particularly when $\alpha$ reaches a high value, such as $\alpha=7$.

\begin{figure}[t] \centering
         \includegraphics[width=\columnwidth,keepaspectratio]{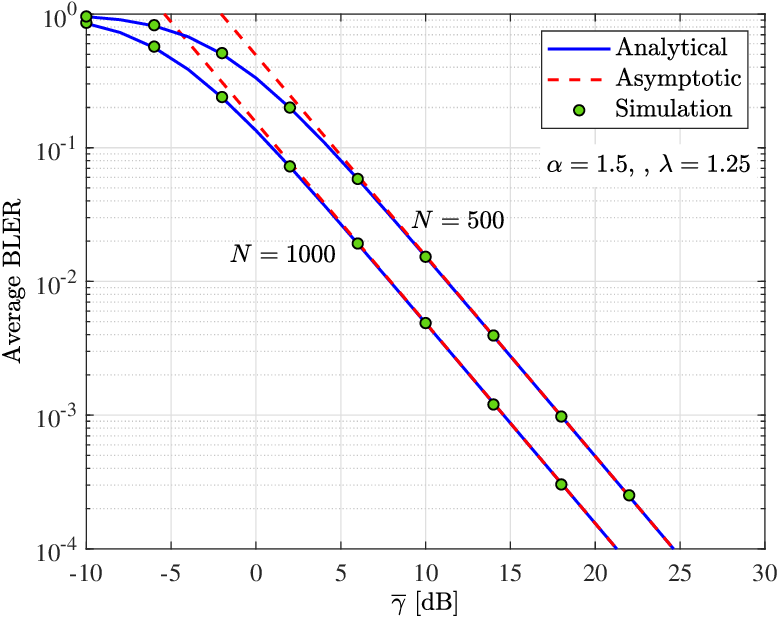}%
       \vspace*{-3mm}
       \caption{Average BLER in short-packet communications.}\label{blerr}
\end{figure}

In Fig. \ref{blerr}, the average BLER performance for short-packet communications shows improvement as the block length increases, as expected. Furthermore, the results validate our analysis's accuracy.
\section{Conclusions}\label{sec5}
In this paper, we introduced a new compound fading channel model known as the $\alpha$-Lomax. In this fading channel model, the reciprocal of the variance
in the Rayleigh distribution is represented by a Gamma RV, and the resulting signal power is not simply obtained as the modulus of the sum of the in-phase and quadrature components. Instead, it is obtained as the modulus raised to a certain given power, represented by the parameter $\alpha$. The $\alpha$-Lomax includes the Lomax fading channel model as a special case, offering more flexibility in modeling wireless fading channels.

\begin{appendix}
\section{}
{\emph{Proof of Lemma 1:}} First, let's derive the signal power PDF of the Lomax distribution{\footnote{Alternatively, it can be  generated when $\mathbb{E}[X^{2}]=\mathbb{E}[Y^{2}]={\tau/2}$, $\tau$ follows the inverse Gamma; that is, $f_{\tau}(\tau) = {\beta^\lambda\over\Gamma(\lambda)}{1\over\tau^{\lambda+1}}\exp{(-{\beta\over\tau})}$ and $f_{P}(p\mid\tau)={1\over\tau}\exp{(-{p\over\tau})}$.}}. To this end, the PDF of \eqref{plomax} can be found using
\begin{align}\label{condpdf}
  f_{P}(p) = \int_{0}^{\infty}f_{P}(p\mid\tau)\,f_{\tau}(\tau)\,\mathrm{d}\tau,
\end{align}
where  $f_{P}(p\mid\tau)=\tau\exp{(-p\,\tau)}$ is the conditional signal power, which follows an exponential RV. Substituting this and \eqref{gamdis} into \eqref{condpdf}, it yields
\begin{align}\label{condpdf1}
\!f_{P}(p) = {\beta^\lambda\over\Gamma(\lambda)}\int_{0}^{\infty} \!\! \,\tau^{\lambda}\exp{(-(\beta+p)\,\tau)}\,\mathrm{d}\tau\stackrel{(a)}{=}{\lambda/\beta\over\left(1+{p\over\beta}\right)^{\lambda+1}},
\end{align}
where $(a)$ is obtained using \cite[Eq. (3.381.4)]{i:ryz}.

The $\alpha$-Lomax fading distribution is defined as $H\triangleq  P^{1\over\alpha}$. Thus, applying transformation of RVs to \eqref{condpdf1}, the PDF of $H$ can be found as
\begin{equation}\label{envaekf}
f_{H}(h) =  {\alpha\lambda\over\beta}\,h^{\alpha-1}\left(1+{h^{\alpha}\over\beta}\right)^{-(\lambda+1)}, \quad y>0.
\end{equation}
For the purpose of channel modeling, let's redefine the PDF in \eqref{envaekf} as follows: Let $Z={H\over\Omega}$, where $\Omega=\mathbb{E}[H]$ denotes the statistical average of $H$, which can be obtained as
\begin{equation}\label{momenvaekf}
\mathbb{E}[H] =  {\beta^{{1\over\alpha}}\Gamma(1+{1\over\alpha})\Gamma(\lambda-{1\over\alpha})\over\Gamma(\lambda)}.
\end{equation}

Using the transformation of RVs and after algebraic manipulations, the PDF, $f_{Z}(z)$, in \eqref{envaekn}, can be readily obtained. Finally, the CDF in \eqref{cdfplomax} can be directly obtained from \eqref{envaekn} through proper integration. This completes the proof of Lemma 1.

{\emph{Proof of Lemma 2:}}
We define the instantaneous SNR as $\Gamma\triangleq{\overline{\gamma}\,Z}$. Thus, one can obtain the desired PDF of the instantaneous SNR, in \eqref{snrpdf}, using $f_{\Gamma}(\gamma)=\frac{1}{\overline{\gamma}}f_{Z}\left({\gamma\over\overline{\gamma}}\right)$. While the CDF, in \eqref{snrcdf}, can be obtained using $F_{\Gamma}(\gamma)=F_{Z}({\gamma\over\overline{\gamma}})$, which ends the proof of Lemma 2.

{\emph{Proof of Lemma 3:}} The generalized MGF can be found by substituting \eqref{snrpdf} into
$
M_{\Gamma}^{n}(s) =  \int_{0}^{\infty}\gamma^{n}\,f_{\Gamma}(\gamma)\exp{(-\gamma\,s)}\,\mathrm{d}\gamma,
$
and sequentially applying \cite[Eq. (8.4.2.5)]{a:pru} and \cite[Eq. (8.3.2.21)]{a:pru}, and using \cite[Eq. (2.25.2.3)]{a:pru}.

{\emph{Proof of Lemma 4:}} The $n$-th moment can be obtained by substituting  \eqref{snrpdf} into
$
\mathbb{E}[\Gamma^{n}] =  \int_{0}^{\infty}\gamma^{n}\,f_{\Gamma}(\gamma)\,\mathrm{d}\gamma,
$
and using \cite[Eq. (3.194.3)]{i:ryz}.
\end{appendix}

\bibliographystyle{IEEEtran}
\bibliography{akefRefe}

\begin{thebibliography}{10}
\providecommand{\url}[1]{#1}
\csname url@samestyle\endcsname
\providecommand{\newblock}{\relax}
\providecommand{\bibinfo}[2]{#2}
\providecommand{\BIBentrySTDinterwordspacing}{\spaceskip=0pt\relax}
\providecommand{\BIBentryALTinterwordstretchfactor}{4}
\providecommand{\BIBentryALTinterwordspacing}{\spaceskip=\fontdimen2\font plus
\BIBentryALTinterwordstretchfactor\fontdimen3\font minus
  \fontdimen4\font\relax}
\providecommand{\BIBforeignlanguage}[2]{{%
\expandafter\ifx\csname l@#1\endcsname\relax
\typeout{** WARNING: IEEEtran.bst: No hyphenation pattern has been}%
\typeout{** loaded for the language `#1'. Using the pattern for}%
\typeout{** the default language instead.}%
\else
\language=\csname l@#1\endcsname
\fi
#2}}
\providecommand{\BIBdecl}{\relax}
\BIBdecl

\bibitem{10309028}
I.~Sanchez and F.~J. Lopez-Martinez, ``The {Lomax} distribution for wireless
  channel modeling: a preliminary study,'' in \emph{Proc. IEEE ECTM}, Ambato,
  Ecuador, Oct. 2023, pp. 1--5.

\bibitem{10356745}
------, ``The {Lomax} distribution for wireless channel modeling: Theory and
  applications,'' \emph{IEEE Open J. Veh. Technol.}, pp. 1--9, Dec. 2023.

\bibitem{Fisher}
S.~K. Yoo, S.~Cotton, P.~Sofotasios, M.~Matthaiou, M.~Valkama, and
  G.~Karagiannidis, ``The {F}isher-{S}nedecor $\mathcal{F}$ distribution: A
  simple and accurate composite fading model,'' \emph{IEEE Commun. Lett.},
  vol.~21, no.~7, pp. 1661--1664, Jul. 2017.

\bibitem{Yacoub2002}
M.~D. Yacoub, ``{The $\alpha$-$\mu$ distribution: a general fading
  distribution},'' in \emph{Proc. IEEE PIMRC}, Sep. 15-18 2002, pp. 629--629.

\bibitem{1622403}
F.~Hansen and F.~Meno, ``Mobile fading—{Rayleigh} and lognormal
  superimposed,'' \emph{IEEE Trans. Veh. Technol.}, vol.~26, no.~4, pp.
  332--335, Nov. 1977.

\bibitem{Abraham}
A.~Abraham, ``Modeling non-{Rayleigh} reverberation,'' SR-266, SACLANT Undersea
  Research Center, La Spezia, Italy, Tech. Rep., 1997.

\bibitem{778479}
A.~Abdi and M.~Kaveh, ``On the utility of gamma {PDF} in modeling shadow fading
  (slow fading),'' in \emph{Proc. IEEE VTC}, vol.~3, Houston, TX, USA, May
  1999, pp. 2308--2312.

\bibitem{9096603}
O.~S. Badarneh, ``The $\alpha$-$\mathcal{F}$ composite fading distribution:
  Statistical characterization and applications,'' \emph{IEEE Trans. Veh.
  Technol.}, vol.~69, no.~8, pp. 8097--8106, Aug. 2020.

\bibitem{i:ryz}
I.~S. Gradshteyn and I.~M. Ryzhik, \emph{Table of Integrals, Series, and
  Products}, 7th~ed.\hskip 1em plus 0.5em minus 0.4em\relax Academic Press,
  California, 2007.

\bibitem{a:pru}
A.~P. Prudnikov, Y.~A. Brychkov, and O.~I. Marichev, \emph{Integrals, and
  Series: More Special Functions}.\hskip 1em plus 0.5em minus 0.4em\relax
  Gordon \& Breach Sci. Publ., New York, 1990, vol.~3.

\bibitem{kilbas}
A.~Kilbas and M.~Saigo, \emph{H-Transforms : Theory and Applications
  (Analytical Method and Special Function)}, 1st~ed.\hskip 1em plus 0.5em minus
  0.4em\relax CRC Press, 2004.

\bibitem{pruv1}
A.~P. Prudnikov, Y.~A. Brychkov, and O.~I. Marichev, \emph{Integrals, and
  Series: More Special Functions}.\hskip 1em plus 0.5em minus 0.4em\relax
  Gordon \& Breach Sci. Publ., New York, 1981, vol.~1.

\bibitem{johnsonsymmetrizing}
D.~Johnson and S.~Sinanovic, ``Symmetrizing the kullback-leibler distance,''
  Rice University, Tech. Rep., 2001.

\bibitem{8270655n}
N.~{Bhargav} \emph{et~al.}, ``On the product of two $\kappa$–$\mu$ random
  variables and its application to double and composite fading channels,''
  \emph{IEEE Trans. Wirel. Commun.}, vol.~17, no.~4, pp. 2457--2470, Apr. 2018.

\end{thebibliography}
\end{document}